\documentclass[ajp,preprint,numbers,unsortedaddress,a4paper]{revtex4-2}

\usepackage[T1]{fontenc}
\usepackage{amsmath}
\usepackage{amssymb}
\usepackage{graphicx}
\usepackage{booktabs}
\usepackage{braket}
\usepackage{xcolor}
\usepackage{listings}
\usepackage{hyperref}
\usepackage{subcaption}

\begin{document}
	\title{The Geometry of Clifford Algorithms: Bernstein-Vazirani as Classical Computation in a Rotated Basis}
	
	\author{Bartosz Chmura}
	\affiliation{Garching (near Munich), Germany}
	\date{\today}
	\email{chmura.quantum@gmail.com}
	
	\begin{abstract}
		The Bernstein-Vazirani (BV) algorithm~\cite{bernstein1997} is frequently taught as a canonical example of quantum parallelism, yet the standard interference-based explanation often obscures its underlying simplicity. We present a geometric reframing in which the Hadamard gate ``wrapping'' acts as a global basis rotation\footnote{keeping in mind requirements of quantum circuits, unitary time evolution, etc.\cite{bernstein1997}.} rather than a generator of computational complexity. This perspective reveals that the algorithm is effectively a classical linear computation over \(GF(2)\) performed in the conjugate Fourier basis, with the apparent parallelism arising from coordinate transformation. 
		
		Building on Mermin's earlier pedagogical shortcut (~\cite[Sec.~2.D.1, Fig~2.10]{mermin2007}), which presented a 'classical' circuit equivalent but stopped short of explicitly labeling it as such, we elevate this to a formal geometric framework. In the extension, we distinguish between globally rotated circuits—which we reveal as classical linear computations—and topologically twisted circuits that generate  quantum entanglement.
		
		We introduce a pedagogical taxonomy distinguishing (1) pure computational-basis circuits, (2) globally rotated circuits (exemplified by Bernstein-Vazirani), and (3) topologically twisted circuits involving non-aligned subsystem bases. This framework allows viewing the Gottesman-Knill theorem (Ref.~\cite[Theorem~1]{gottesman1998}) from a new angle, extends students' understanding of phase kickback and the 'Ricochet Property'~\cite{larocca2022, preskill2016, wilde2017}. Furthermore it provides a more intuitive starting point for explaining Bell-pair extensions through concrete circuit derivations and Qiskit simulations suitable for undergraduate quantum information courses. The outlook explores how this geometric view paves the way for understanding entanglement as topological twists.
	\end{abstract}
	
	\maketitle
	
	\section{Introduction: Clarifying the Origins of Parallelism}
	
	In the pedagogy of Quantum Information Science (QIS), the Bernstein-Vazirani (BV) algorithm~\cite{bernstein1997} occupies a unique position. It is often the first algorithm introduced that demonstrates a deterministic, polynomial separation between quantum and classical query complexity in the oracle model, serving as a foundation for super-polynomial extensions in recursive Fourier sampling~\cite{aaronson2004}. The standard narrative—centered on superposition enabling ``evaluation on all inputs simultaneously'' followed by destructive interference (or uncomputing)—frequently leaves students puzzled about the true source of the advantage~\cite{mermin2007}.
	
	This paper proposes an alternative geometric view: the Hadamard gate is not primarily a ``mixer'' creating complexity, but a rotation operator reorienting the circuit components between the computational and Fourier bases (as in Chapter 3.3.4 in~\cite{wilde2017}), thereby aligning the computational frame with the oracle's structure. In this perspective, BV belongs to a family of globally rotated Clifford circuits that remain classically simulable, as established by the Gottesman-Knill theorem (Ref.~\cite[Theorem~1]{gottesman1998}, Ref.~\cite[Theorem~2]{vandennest2010}), revealing ``quantum parallelism'' as a basis-dependent pheonomenon rather than an indispensable resource.
	
	This work emphasizes hands-on step-by-step circuit transformations and simulations as first steps in building geometrical intuition behind the mathematical transformations supporting the Gottesman-Knill theorem~\cite{gottesman1998}. 
	The approach serves a two-pronged strategy: explicitly visualizing transformations while accustoming students to rigorous verification—prompting questions like 'Why do we need to verify?', 'Where could it fail?', and 'What aspects of our intuition may mislead us?'. These are relevant both in the sense of assuring 'Is our circuit transformation valid?' and in the perspective of adopting a recurrential approach as mentioned in~\cite{aaronson2004}.
	Furthermore, it introduces the concepts of ``geometric rotations'' and ``geometric twists'' in circuit design, preparing the ground for advanced discussions on how non-aligned bases between circuit components generate entanglement.

	\section{The Standard Derivation: Interference Narrative}
	
	The BV problem (referred to as the "parity problem" in Ref.~\cite[Sec.~8.4]{bernstein1997}) asks a computer to identify a secret string \( s \in \{0,1\}^n \) hidden inside a function \( f(x) = s \cdot x \pmod 2 \).
	
	\begin{itemize}
		\item \textbf{Classically:} \( n \) queries are required to reveal the string bit-by-bit.
		\item \textbf{Quantumly:} One query suffices.~\cite{aaronson2004}
	\end{itemize}
	
	The standard circuit proceeds as follows:
	\begin{enumerate}
		\item Initialize \( \ket{0}^{\otimes n} \ket{1} \) (ancilla for phase kickback).
		\item Apply \( H^{\otimes n+1} \) (creating \( \ket{-} \) on the ancilla).
		\item Oracle \( U_f \) (Standard Phase Oracle).
		\item Final \( H^{\otimes n} \) on the query register.
		\item Measure: obtain \( s \) deterministically.
	\end{enumerate}
	
	\textbf{Mechanism:} The standard explanation (Ref.~\cite[Chap.~2.D.1]{mermin2007}) relies on Phase Kickback encoding \( (-1)^{f(x)} \) into the amplitudes, followed by a final Hadamard layer performing an inverse Quantum Fourier Transform over \( \mathbb{Z}_2^n \).
	
	\section{The Geometric Derivation: Global Basis Rotation}
	
	\subsection{The Fourier Basis as Orthogonal Coordinates}
	
	We treat the computational ($Z$) and superposition ($X$/Fourier) bases democratically as coordinate systems related by rotation, which aligns with Mermin’s approach (Ref.~\cite[Chap.~1]{mermin2007}).
	We define the Fourier basis states as:
	\[
	\ket{0}_X \equiv \ket{+} = \frac{\ket{0} + \ket{1}}{\sqrt{2}}, \quad \ket{1}_X \equiv \ket{-} = \frac{\ket{0} - \ket{1}}{\sqrt{2}}
	\]
	The Hadamard gate \( H \) is the rotation operator \( Z \leftrightarrow X \), consistent with the Heisenberg representation of Clifford gates (Ref.~\cite[Table~1]{gottesman1998}).
	The fundamental identity of the Clifford group required for this analysis is:
	\begin{equation}
		H Z H = X.
	\end{equation}
	This identity states that a Phase Flip ($Z$) in the computational frame is mathematically identical to a Bit Flip ($X$) in the Fourier frame.
		
	\subsection{Conjugating the Circuit}
	
	While the basis-transformation of the BV oracle has been noted as a simplified pedagogical shortcut (Ref.~\cite{mermin2007}, Sec.~2.D.1, Figs. 2.8-2.10), we elevate this observation into a formal geometric framework. 
	
	Mermin provided an equivalent 'classical' circuit but, perhaps to preserve quantum intuition, shied from explicitly putting it as such in the spotlight—we do so here, framing it as classical linear computation in the rotated basis.
	
	By emphasizing the Hadamard `wrapper' as a basis rotation applied to the internal circuit rather than a generator of computational complexity, we reveal the algorithm's true identity: a classical linear computation over GF(2) evaluated in the conjugate Fourier basis.
	
	Furthermore, because these geometric coordinate transformations are unitary and symmetric, the derivation is entirely bidirectional; the student can derive the ``standard'' quantum BV circuit from the ``classical'' parity circuit, or vice versa.
	
	Pedagogically, this requires providing the student two perspectives of viewing a quantum circuit (or its parts): 
	- as a chronological sequence of discrete time-steps,
	- and analyzing it as a single, geometric object. 
	This spatial, diagrammatic approach shares a deep conceptual affinity with the ZX-calculus~\cite{vandewetering2020}, an advanced graphical language where Hadamard wrappers (H-boxes) seamlessly alter the topology of quantum processes. By introducing this geometric reasoning early using standard circuit barriers to delimit the operational blocks, educators can provide students with a highly intuitive stepping stone toward modern diagrammatic formalisms. 
		
	The authors of this work believe that it is best left in the educator's discretion to decide which directions to introduce in which order to their students.
	
	Our choice, for the most striking pedagogical effect and possibly least difficult intuition building, is to start with the classical circuit - taking to heart Mermin's ``much easier way to understand what is going on'' comment. 
	
	\subsubsection{Overview of the Geometric Transformation}
	Before detailing the step-by-step circuit equivalences, it is helpful to outline the overall geometric strategy. The following sequence of diagrams demonstrates a continuous structural deformation from the standard quantum Bernstein-Vazirani algorithm (with its apparent interference and superposition) into a purely classical linear computation over GF(2). 
	
	Rather than deconstructing the standard quantum circuit only as a chronological sequence of time-steps, we will build it in reverse:
	\begin{itemize}
		\item Fig~\ref{fig:classical_view} - the classical algorithm starting point,
		\item Fig~\ref{fig:step2_cz} - $CX / CZ$ transformation is used,
		\item Fig~\ref{fig:step3_h_move} - $H$ gates are moved out,
		\item Fig~\ref{fig:step4_sandwich} - remaining $H$ are introduced from $I=HH$,
		\item Fig~\ref{fig:ancilla_h} - a final set of $CZ / CX$ transformations is applied,
		\item Fig~\ref{fig:final_bv} - $HH=I$ gates are simplified, target BV form is achieved.
	\end{itemize} 
	This visual proof confirms that the computational complexity does not stem from quantum parallelism, but merely from the coordinate system in which the classical logic is embedded.
	
	\subsubsection{A Note on Circuit Depth and Parallel Execution:} While the step-by-step geometric transformations in the accompanying figures (such as the expanding Hadamard layers in Fig.~\ref{fig:step4_sandwich} and Fig.~\ref{fig:ancilla_h}) are drawn sequentially to emphasize structural changes, this does not imply a strictly sequential execution or growing circuit depth in actual physical hardware. 
	
	While the canonical BV circuit introduces $O(1)$ depth overhead due to the global $H^{\otimes n}$ basis rotations, our geometric reduction to the Z-basis reveals that this is purely an artifact of a coordinate mismatch. 
	While the standard algorithmic definition of quantum parallelism—evaluating a function over a coherent superposition of computational-basis states—remains valid for the standard BV circuit, our reduction demonstrates that this parallelism can be entirely bypassed via geometric transformations. 
	Because the identical $O(1)$ query complexity is naturally recovered in the conjugate Fourier basis (which contains no superposition), educators can offer a concrete, parallel-free alternative: one can either evaluate a superposition of inputs using phase kickback and interference in the Z-basis, or treat the initial and final Hadamards as transforming the oracle itself into a classical operation in the X-basis. 
	Ultimately, whether one observes simultaneous evaluation or a classical sequential query depends entirely on the chosen coordinate lens. The physical depth of the computation remains strictly the depth of the classical $CZ/CX$ oracle.
	
	Nevertheless, this allows us to make two great didactic points:
	\begin{itemize}
		\item the circuit depth may vary dependent on the actual hardware architecture, which may be directly shown by the circuit geometrical transformations,
		\item the circuit irreducible form depends on its topological properties: while the BV algorithm circuits (Family II) can be reduced to a classical algorithm, others like Deutsch-Jozsa (Family II/III - see Appendix A discussion) may not be reducible to such form anymore. 
	\end{itemize}

	\subsubsection{Transformation Details}	
\begin{figure}[h]
		\centering
		\includegraphics[width=\columnwidth]{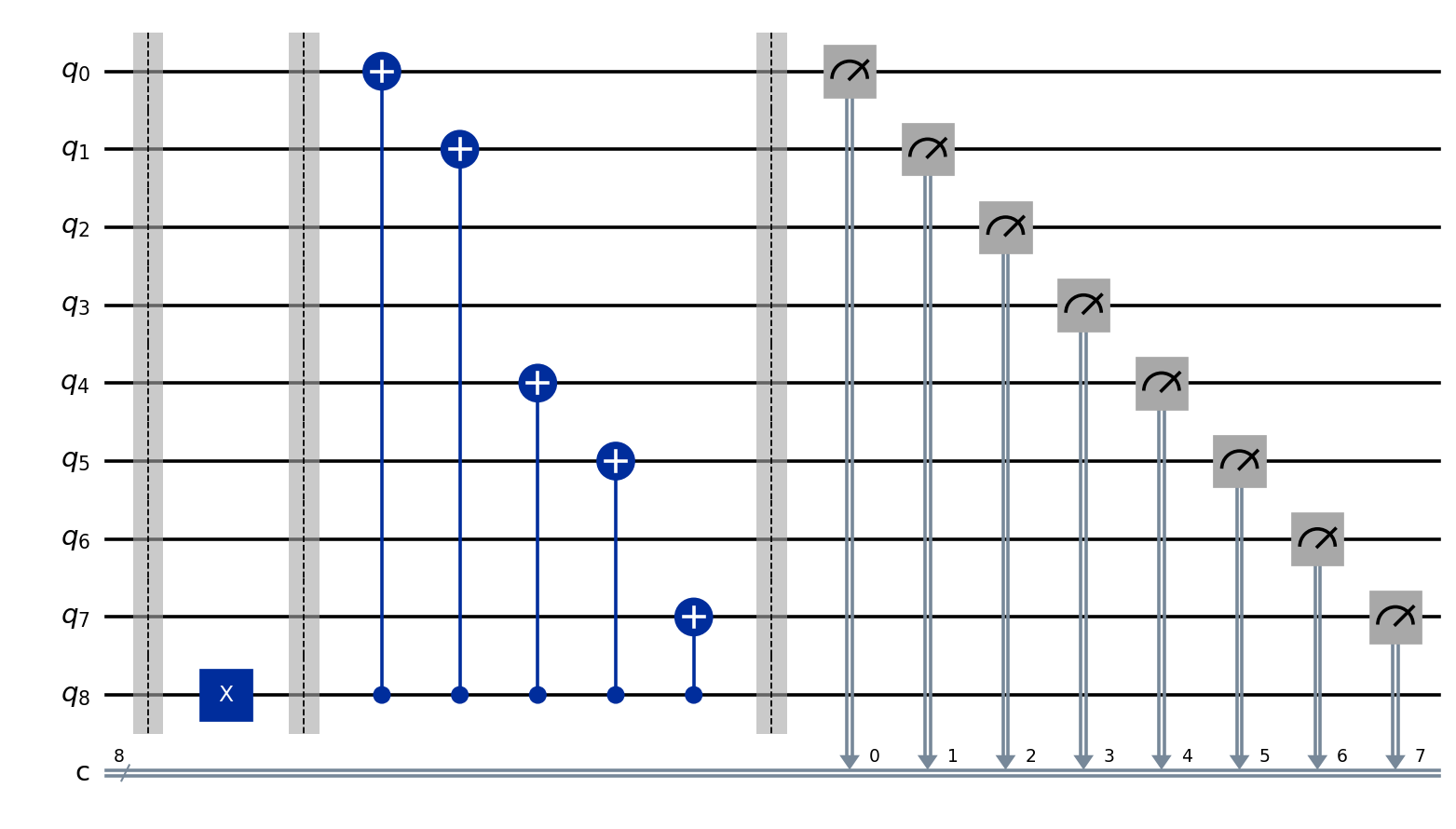}
		\caption{The ``Classical'' view of the Bernstein-Vazirani circuit. When viewed in the COMPUTATIONAL basis, the oracle acts simply as a set of CNOT gates controlled by the ancilla ($q_8$), effectively writing the secret string directly onto the data wires. The barriers clearly delimit the oracle operation from the initialization and measurement steps. The ``Canonical'' Fourier basis view (Hadamard gates wrapper) is currently hidden in the CNOT gates.}
		\label{fig:classical_view}
	\end{figure}

	We begin with a deterministic GF(2) parity operation, a classical circuit as shown in Fig.~\ref{fig:classical_view}. The circuit conditionally writes a string into the measurement apparatus. The ancilla qubit via X-gate simply provides the circuit with a 'write' instruction. Code:~\ref{lst:circuit_1_bv_xwrite}.
	We delimit this core operation with barriers to encourage analyzing the sub-circuit as a single, holistic geometric object—a spatial perspective that allows introduction of advanced diagrammatic reasoning.
	
	Notice: one could further simplify the circuit by replacing $CX$ gates with $X$ and completely removing the ancilla - this will become relevant when taking into account the Appendix A and discussing the Deutsch-Jozsa example.
	
\begin{figure}[h]
		\centering
		\includegraphics[width=\columnwidth]{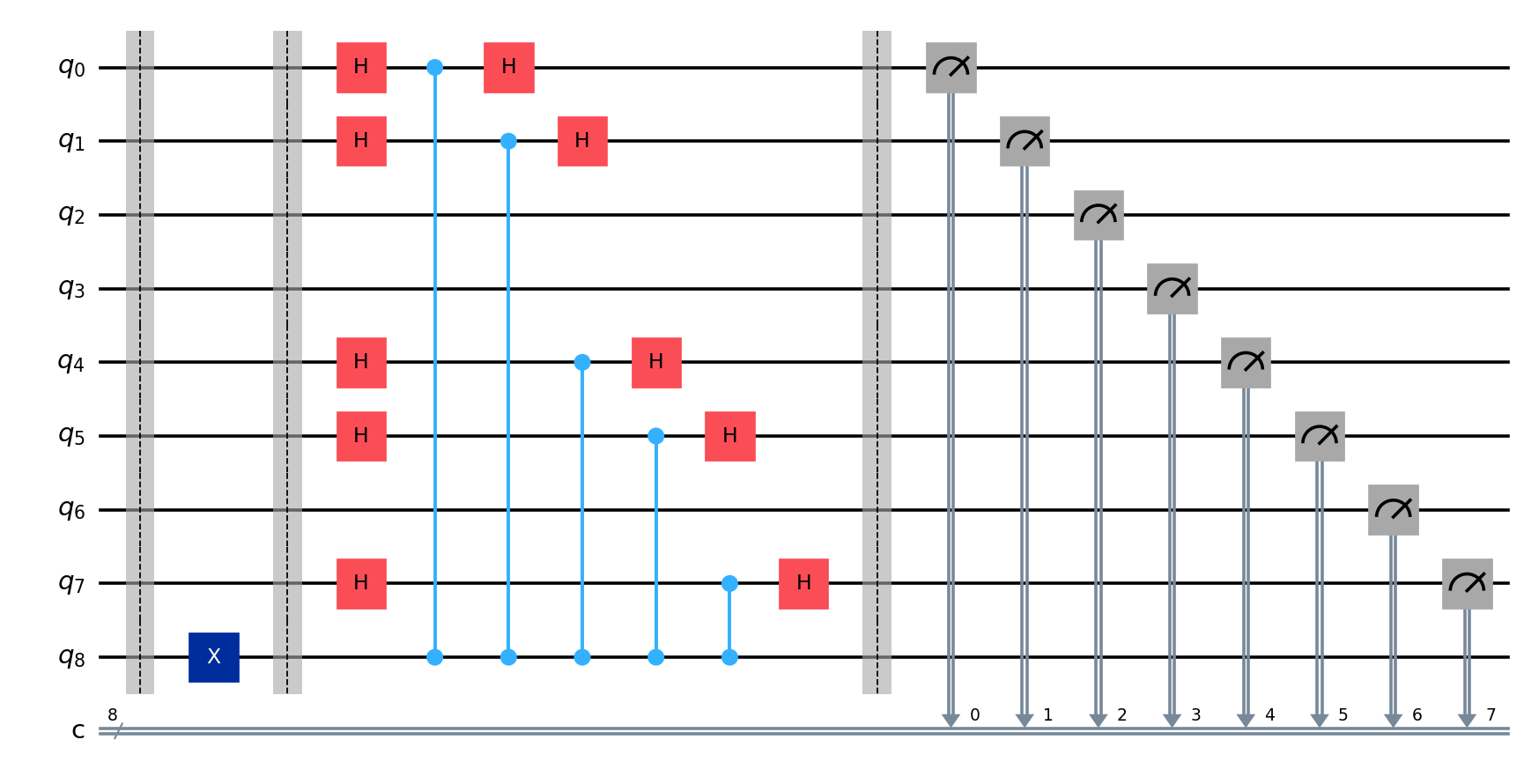}
		\caption{Intermediate geometric transformation. We replace the CNOT gates (from Fig.~\ref{fig:classical_view}) with their Clifford equivalent: a Phase-Controlled ($CZ$) gate sandwiched by Hadamards on the target qubits. Physically, we have rotated the data qubits from the bit-flip axis ($X$) to the phase-flip axis ($Z$), enabling the interaction to be described as a phase query.}
		\label{fig:step2_cz}
	\end{figure}

	The next step is to use the \( H X H = Z \) identity and transform the \(CX\) gates into \( CZ \) resulting in the circuit as presented in Fig.~\ref{fig:step2_cz}. Code: \ref{lst:circuit_2_bv_zcz}.
	
\begin{figure}[h]
		\centering
		\includegraphics[width=\columnwidth]{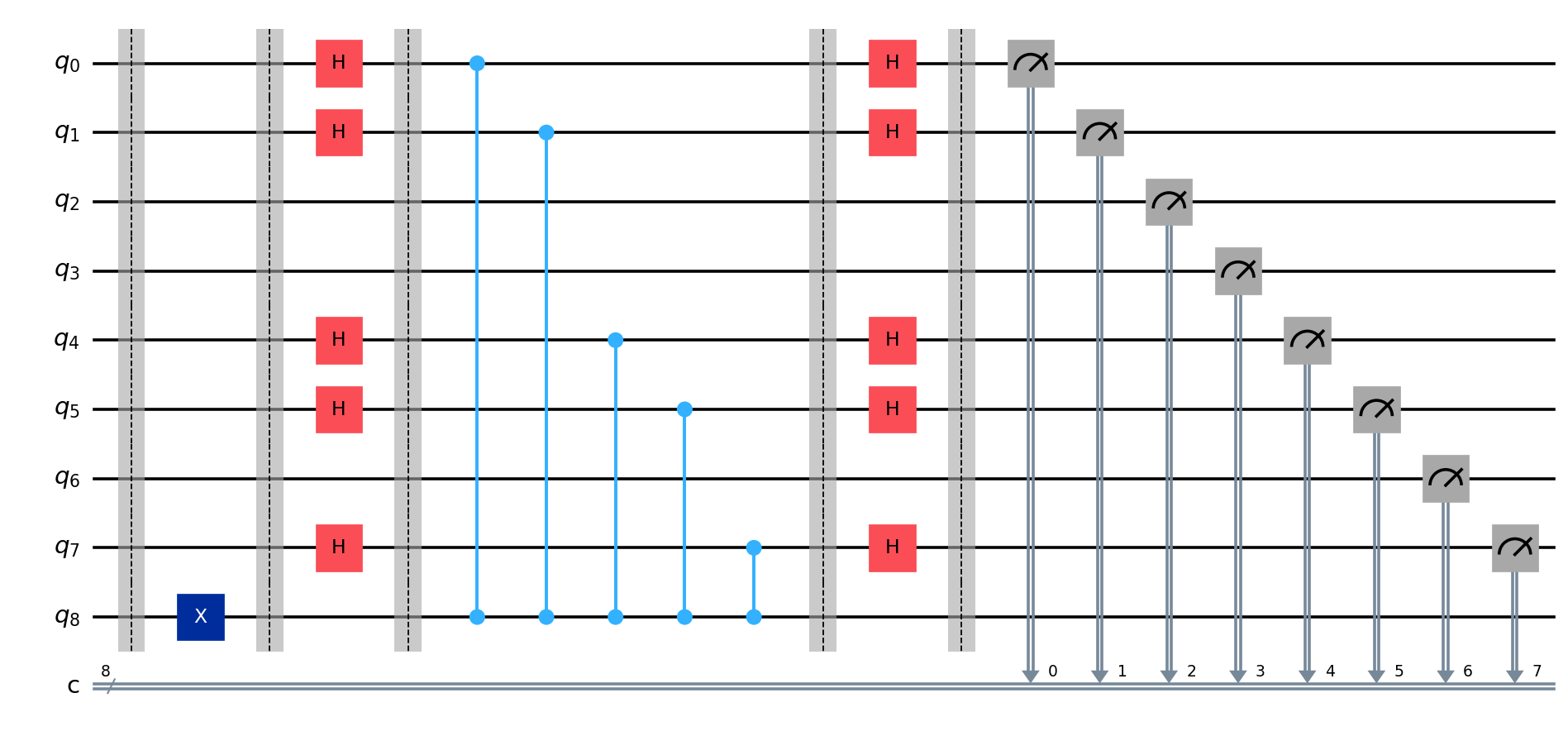}
		\caption{Structural rearrangement. We slide the inner Hadamard gates to the boundaries of the oracle block (marked by barriers). This visual grouping highlights the ``Hadamard Sandwich'' structure: the central component is a pure Phase Oracle ($CZ$), and the surrounding Hadamards serve as the coordinate transformations that rotate the basis.}
		\label{fig:step3_h_move}
	\end{figure}

	Then, in order to prepare the suggested geometry-first approach, we move the Hadamard gates to the edge of the oracle as shown in Fig.~\ref{fig:step3_h_move}, so the circuit-component-wise rotation can become apparent.
	
\begin{figure}[h]
		\centering
		\includegraphics[width=\columnwidth]{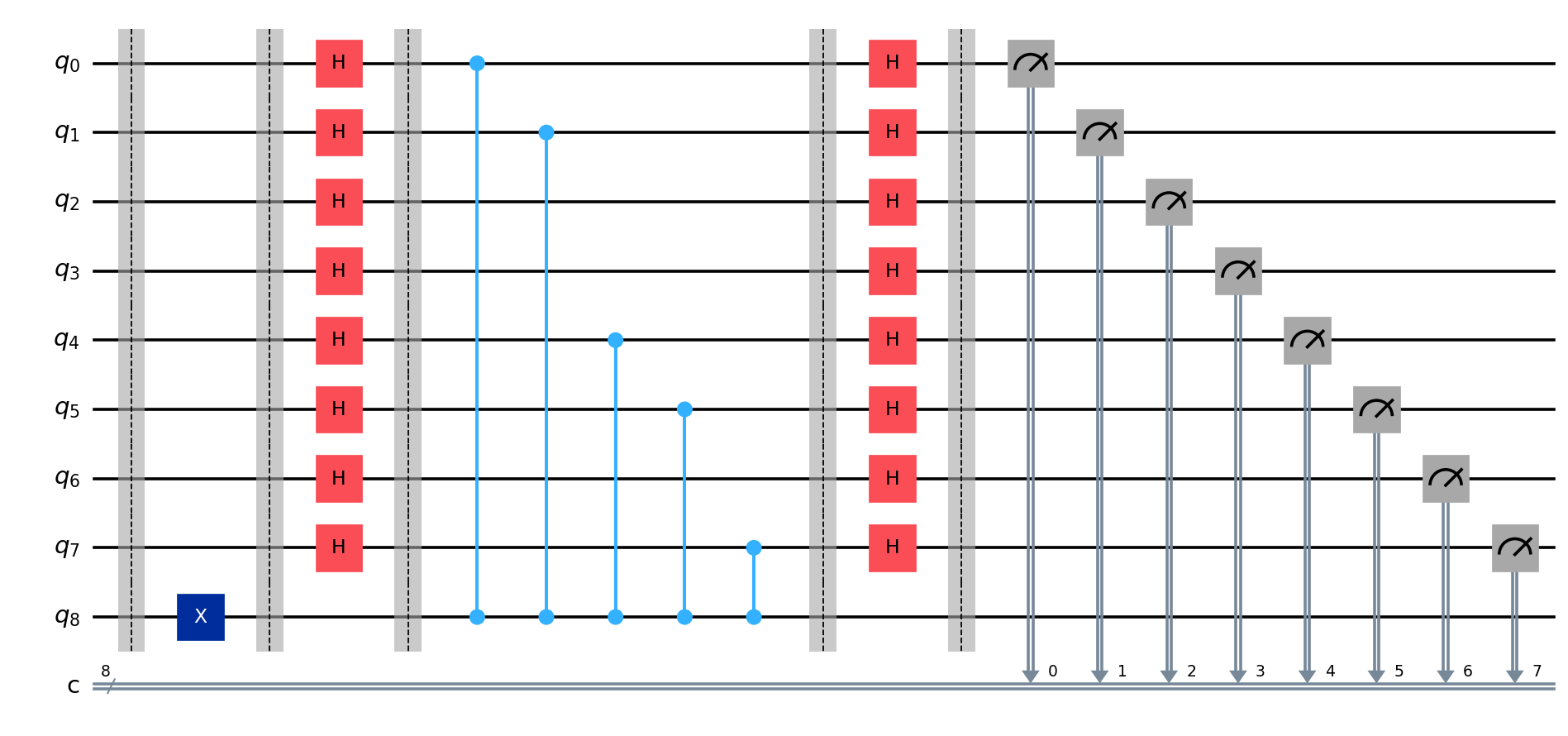}
		\caption{The Global Rotation (Standard BV Circuit). By completing the Hadamard layers on all oracle qubits (using $I=HH$), we arrive at the phase-representation of the Bernstein-Vazirani algorithm. Here, the oracle is defined in the $Z$-basis (Phase Oracle) and the Hadamard gates act as the global basis change. This derivation demonstrates that the ``Quantum'' circuit is merely the ``Classical'' circuit (Fig.~\ref{fig:classical_view}) viewed from the Fourier basis.}
		\label{fig:step4_sandwich}
	\end{figure}

	Completing this part of transformation rests on applying the \( I = H H \) identity on the qubits unaffected by the \( CZ \) gates. As a result we obtain the first result, where the oracle is clearly visible as a ($CZ$) ``sandwiched'' by Hadamard transformations - Fig.~\ref{fig:step4_sandwich}.
	
\begin{figure}[h]
		\centering
		\includegraphics[width=\columnwidth]{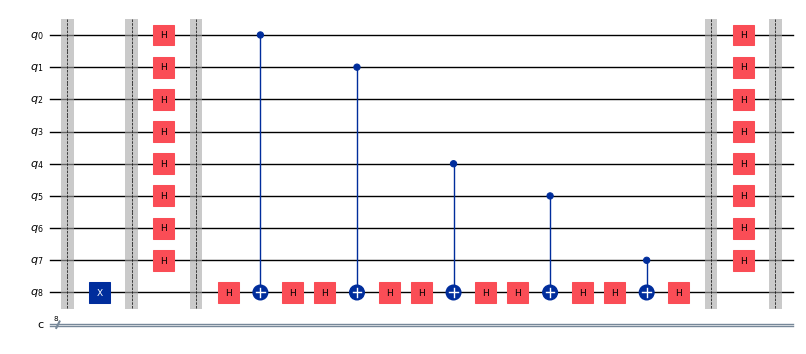}
		\caption{The Phase Kickback Construction. Here we expand the internal structure of the Phase Oracle. By decomposing the $CZ$ gates into $CNOT$ gates sandwiched by Hadamards on the target (ancilla), we recover the standard experimental implementation of Bernstein-Vazirani. The ancilla is effectively prepared in the $|-\rangle$ state (via the $X$ and $H$ gates), causing the bit-flips triggered by the CNOTs to ``kick back'' a phase of $-1$ to the control qubits.}
		\label{fig:ancilla_h}
	\end{figure}

	The step before last in recovering the standard textbook implementation is to apply the \( Z = H X H \) identity specifically to the target (ancilla) qubit of each gate. We transform the symmetric \( CZ \) gates into \( CNOT \) gates where the data qubits act as controls and the ancilla acts as the target. As a result, the oracle is revealed to be a sequence of CNOTs sandwiched by Hadamard gates on the ancilla line. This explicitly visualizes the ``Phase Kickback'' mechanism: the bit-flip ($X$) on the ancilla is converted into a phase-flip ($Z$) on the control by the surrounding basis rotations, as shown in Fig.~\ref{fig:ancilla_h}.
	
\begin{figure}[h]
		\centering
		\includegraphics[width=\columnwidth]{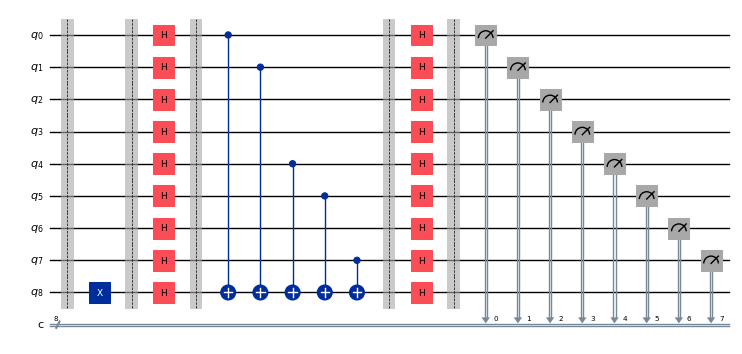}
		\caption{The Canonical Bernstein-Vazirani Circuit. This is the standard form found in introductory textbooks. All qubits start in $|0\rangle$, pass through a global Hadamard layer (creating superposition), interact via a CNOT-based oracle (utilizing phase kickback from the $|-\rangle$ ancilla), and undergo a final Hadamard transformation before measurement. The geometric derivation agrees with the Fourier transform of a classical bit-write operation.}
		\label{fig:final_bv}
	\end{figure}

	The final step involves consolidating the Hadamard transformations on the ancilla line. By applying the identity \( I = H H \) and grouping the remaining gates to the exterior, we recover the complete, canonical textbook representation of the Bernstein-Vazirani algorithm. This concludes the derivation, proving that the standard quantum circuit is mathematically isomorphic to the classical circuit shown in Fig.~\ref{fig:final_bv}, connected simply by a change of coordinate basis. Code: \ref{lst:circuit_3_bv_zcx}.
	
	This approach lets us look at the BV circuit not as a sequence of time-steps, but as a single geometric object. The circuit consists of an oracle \( U_f \) ``sandwiched'' by Hadamard transformations: \( H^{\otimes n} U_f H^{\otimes n} \).
	
	\textbf{Taking the reverse path}: If we ``slide'' the Hadamard gates through the oracle (using the identity \( H Z H = X \)), the oracle's structure changes. The conjugation transforms the interaction:
	\begin{equation}
	(H \otimes I) \cdot \text{CNOT}_{c,t} \cdot (H \otimes I) = (H \otimes H) \cdot \text{CZ}_{c,t} \cdot (H \otimes H) = (I \otimes H) \cdot \text{CNOT}_{t,c} \cdot (I \otimes H)
	\label{eq:cx-cz-cx-1}
	\end{equation}
	Note the reversal of control and target.
	
	\textbf{Result:} In the Fourier frame, the input \( \ket{0}^{\otimes n} \) maps to the \( \ket{+}^{\otimes n} \) state (logical zero in rotated coordinates). The ``quantum'' oracle transforms into a classical array of CNOT gates that mechanically write \( s \) onto the wires. There is no search; there is simply a write operation viewed from a rotated angle.
	
	\section{A Taxonomy of Circuit Complexity}
	
	The geometric perspective developed here motivates a classification of Clifford-group circuits based not on their operation counts, but on their \textit{basis alignment}. We propose three distinct families:
	
	\subsection{Family I: Pure Z-Basis (Classical)}
	Circuits constructed entirely using Computational Basis gates (e.g., $X$, $CNOT$, Toffoli). These are standard reversible classical logic circuits. There is no superposition, and the system remains in a product state relative to the $Z$-basis throughout the computation.
	
	\subsection{Family II: Globally Rotated (Hidden Classical)}
	Circuits where the entire register (or independent subsystems) can be described by a global basis rotation that simplifies the operation back to Family I  (Ref.~\cite[Sec.~6]{vandennest2010}). The Bernstein-Vazirani algorithm is the archetype of this family. While the standard "parity problem" is efficiently solvable quantumly with a single query, it does serve as the base case for recursive structures where "uncomputing requirements" impose fundamental limits~\cite{aaronson2004} - or to put simply, many 'Family II' subsystems may be used to construct a 'Family III' type system. 
	In the BV case - while the state vector appears complex in the $Z$-basis, the intrinsic logic is classical linear algebra in the Fourier basis.
	
	\textbf{Pedagogical Bridge: Parity Gates and the Ricochet Property} 
	Family II serves as the perfect setting to de-mystify "dynamic" quantum effects by revealing them as static geometric symmetries. 
	\begin{itemize}
		\item \textbf{Phase Kickback as Parity Rotation:}  We show an alternative perspective of "phase kickback" - not as a directional flow of information, but as a basis-dependent view of symmetric parity gates. 
		
		Starting from a $ZZ$-parity gate (physically equivalent to a $CZ$ up to single-qubit phases), we  can show the student an alternative formulation of eq.\ref{eq:cx-cz-cx-1} in which using the $HZH=X$ transformation rotates the interaction frame: $XZ \leftrightarrow ZZ \leftrightarrow ZX$. 
		To this extent we suggest making a tangent focusing on the $CX-CZ$ transformations when analyzing: Fig.\ref{fig:classical_view}-Fig.\ref{fig:step2_cz}-Fig.\ref{fig:ancilla_h}.
	
		In this picture the "kickback" is shown as an effect of transformations of the $ZZ$ parity gate viewed through a locally-rotated (twisted) frame (keeping in mind: $H=H^T$):
		\begin{equation}
			(I \otimes H) \cdot ZX \cdot (I \otimes H) = ZZ = (H^T \otimes I) \cdot XZ \cdot (H^T \otimes I) = (H^T \otimes H) \cdot XX \cdot (H^T \otimes H)
			\label{eq:zx-zz-xz-xx}
		\end{equation}
		
		\item \textbf{The Ricochet Property:} Extending the last transformation allows a smooth geometric introduction of the "Ricochet Property" or "transpose trick"—a term used in recent literature to describe the geometric symmetry where operations on one qubit are equivalent to transposed operations on a coupled partner~\cite{larocca2022, preskill2016, wilde2017}. 
		While the simple gate transformation promoted to a circuit-wide rotation in Family II, in case of the Ricochet property, serves a perfect precursor to the "Topological Twists" of Family III.
		
		\item \textbf{Connection to Choi-Jamiolkowski isomorphism:} Pedagogically, analyzing these symmetries provides an accessible entry point to more advanced topics like the Choi-Jamiolkowski isomorphism (CJI) found in foundational information theory~\cite{nielsen2000}. By observing how a gate (channel) relates to the basis alignment (state), students can begin to grasp the duality between quantum operations and entangled states with additional, geometric intuition, to support the matrix formalism. Introducing the geometric notion of 'bending wires' is much easier to accept, when introduced together with, or after other geometric concepts of 'wire twists' performed by single-qubit gates: $H, S, X, Y, Z$. A simple example for pedagogy could employ an adapted 'belt / Dirac-string trick' as used in visualizing behavior of spins (as shown in \cite{staley2010}) - where students could be shown the 'transpose trick' (see Exercise 3.7.12 in \cite{wilde2017}) as moving the belt twist between its ends representing the Bell pair 'bent wire'.
	\end{itemize}
	
	\subsection{Family III: Topologically Twisted}
	
	Circuits where subsystems are rotated into \textit{non-commuting} bases relative to each other. A canonical example is Bell State preparation, where one qubit acts in the $X$-basis while its partner acts in the $Z$-basis. 
	
	Unlike Family I and II circuits, which merely rotate separable states within local geometric bases, for Family III no single global coordinate frame exists that can reduce the state to a simple product. 
	This ``Topological Twist''—the misalignment of local frames—is the geometric origin of entanglement. Introducing this family opens a fascinating pedagogical door to investigating Bell- and Cluster States.
	The circuits of this family apply non-linear phase polynomials (the simplest being the $CZ$ gate) that generate true bipartite or multipartite entanglement.
	These circuits can be classified as ``genuinely quantum,'' as they force the computational register into entangled subspaces characterized by non-trivial topological and geometric structures that cannot be factored into local operations~\cite{bengtsson2017, mosseri2001}.
	However the Gottesman-Knill theorem allows a further breakdown into Clifford and post-Clifford Family III subgroups, but a detailed discussion exceeds the frame of this paper.
		
	\section{Circuit Simulations in Qiskit}
	
	To validate this geometric reframing, we implement the circuit transformations in Qiskit. We demonstrate that the standard BV circuit produces identical statistics to the ``classical'' circuit in the rotated basis. The complete Python code (in Jupyter Notebook format) used for these simulations is provided in Appendix B.
	
	The Jupyter Notebook containing the full executable code is also available as an ancillary file accompanying this preprint on arXiv.

	\section{Limitations and Future Work}
	
	While the geometric reframing presented in this paper offers a rigorous circuit equivalence, it is primarily proposed as a theoretical pedagogical framework. 
	Its practical efficacy in preventing student misconceptions compared to the standard interference-based narrative remains to be verified through formal Physics Education Research (PER) in a classroom setting. 
	Furthermore, this approach does not attempt to resolve fundamental ontological debates regarding the physical reality of superposition; it merely utilizes basis transformations to optimize the representation of the circuit for teaching purposes.
	
	Future work will naturally extend this geometric taxonomy beyond the globally rotated Family II circuits. A forthcoming paper will explore how algorithms introducing non-linear boolean terms—such as the quadratic Deutsch-Jozsa oracle—geometrically fracture the Pauli normalizer. Analyzing these topologies will visually demonstrate the transition from classically simulable Clifford entanglement to universal quantum computation.
	
	Furthermore, the taxonomy introduced here suggests a deeper geometric link. We propose that quantum entanglement can be rigorously understood as a ``twist'' in the interaction geometry of the circuit. Future work will explore how these local twists in the Pauli group generate the non-local correlations characteristic of cluster- and Bell states.
	
	\section{Conclusion}
	
	By recasting the Bernstein-Vazirani algorithm as a classical computation in a rotated basis, we provide a visual framework that allows educators and students to view ``quantum parallelism'' as a basis-dependent phenomenon. 
	Seen through the lens of the Fourier basis, the algorithm's apparent ``exhaustive search'' over a superposition transforms into a deterministic write operation after proper frame alignment.
	
	This pedagogical reframing serves as a launchpad for more advanced concepts. At the very least, it suggests that quantum algorithms may be constructed by starting with a classical circuit in the computational basis and rotating it into another basis (a Family II construction).
	
	While this paper's main focus has been on the algorithmic and structural deconstruction of the Bernstein-Vazirani circuit, the physical implications of this perspective are equally profound. Reinterpreting the oracle's operation through the lens of continuous gauge transformations and fractional translations offers a deeper physical intuition into why such circuits are classically simulable. To emphasize the consequences of this approach, in Appendix A we provide an example of deriving the irreducible form of a Deutsch-Jozsa circuit (which itself sits on the boundary between Family II and Family III circuits).
	
	\section*{Appendix A: Z-Basis Reduction of a Quadratic Deutsch-Jozsa Oracle as an example of a Family III circuit}
	\label{app:dj_reduction}
	
	While the Bernstein-Vazirani (BV) algorithm evaluates strictly linear Boolean functions and serves as the archetype for Family II (Globally Rotated) circuits, the Deutsch-Jozsa (DJ) algorithm relaxes this constraint to allow any balanced Boolean function. When a balanced function contains non-linear terms, its phase-basis representation requires entangling operations, pushing the circuit into Family III (Topologically Twisted).
	
	To demonstrate an example application of the geometric rotation framework for a Family III circuit, we provide a step-by-step gate-level reduction of a quadratic DJ oracle.
	
	\subsection{The Standard View: Complex Entanglement}
	Consider a 4-qubit DJ oracle designed to mark exactly four states (e.g., $|011\rangle$, $|100\rangle$, $|101\rangle$, and $|110\rangle$ - in Qiskit notation). 
	In the standard computational basis, this oracle is constructed using a sequence of Multi-Controlled-X (Toffoli) gates wrapped in $X$ gates to trigger on specific bit-patterns. 
	
	\begin{figure}[h]
		\centering
		\begin{subfigure}[b]{0.48\linewidth}
			\centering
			\includegraphics[width=\linewidth]{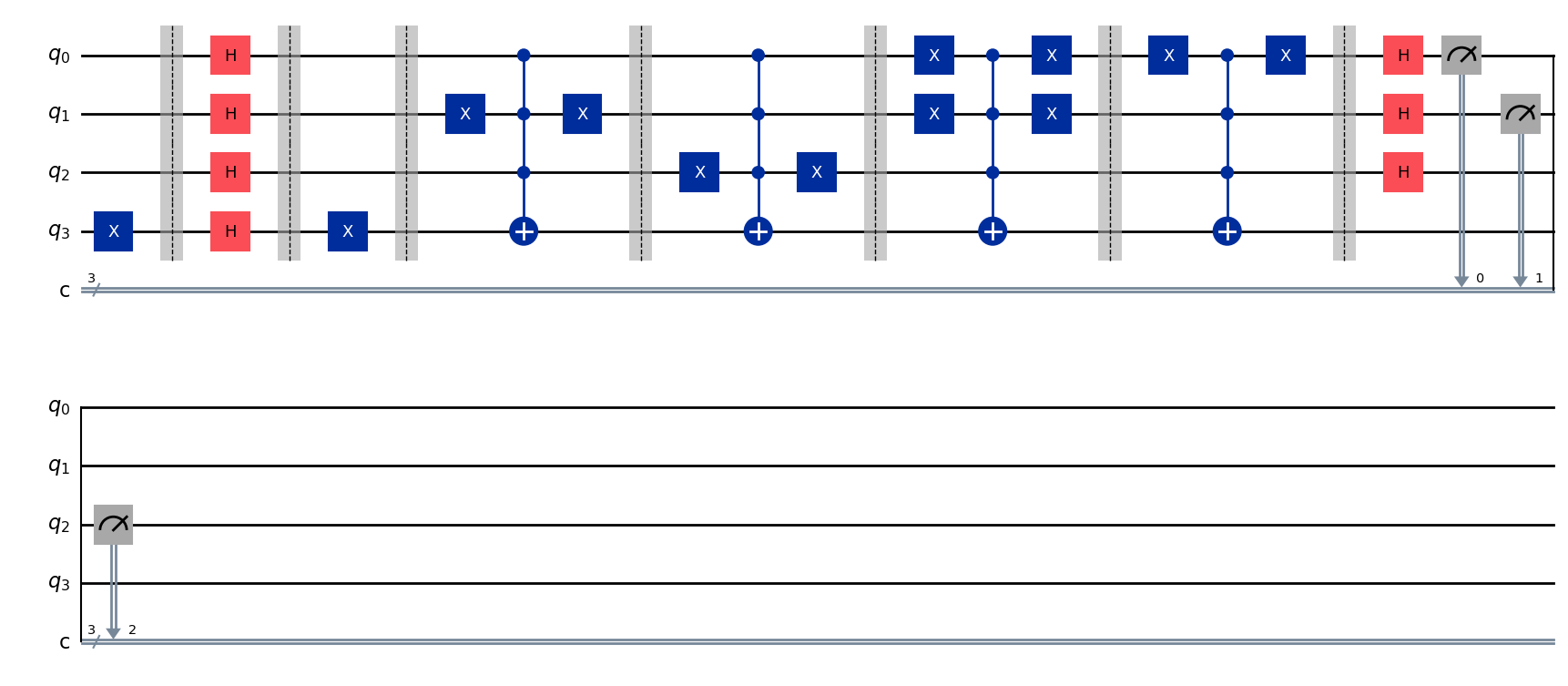}
			\caption{Initial representation}
			\label{fig:dj_standard_a}
		\end{subfigure}
		\hfill
		\begin{subfigure}[b]{0.48\linewidth}
			\centering
			\includegraphics[width=\linewidth]{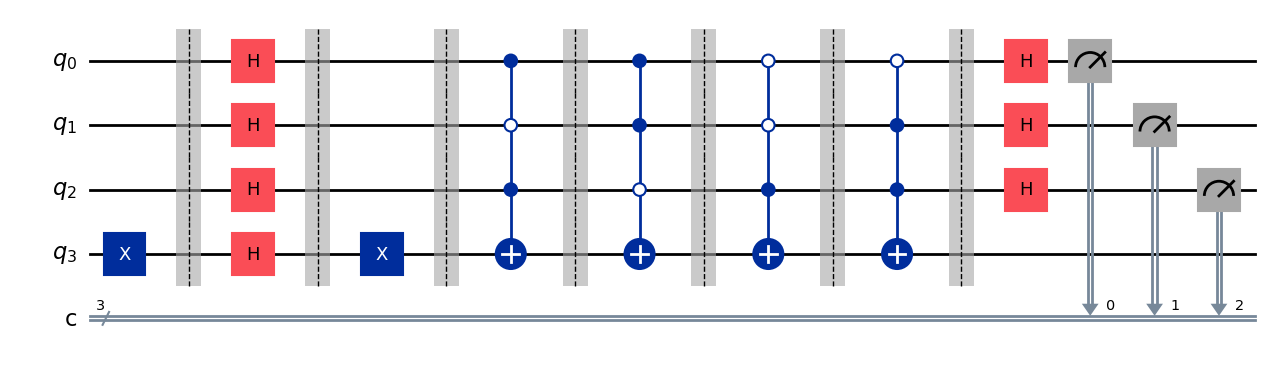}
			\caption{Alternative representation}
			\label{fig:dj_standard_b}
		\end{subfigure}
		
		\caption{The DJ algorithm circuit in two representations, a) using MCX gates and b) Multi-Controlled-Open-X gates.}
		\label{fig:dj_standard}
	\end{figure}
	
	In this basis, the circuit appears highly complex, requiring multi-qubit physical bit-flips conditioned on the entire register state. Another, visually more compelling representation of the same circuit shown on Fig.~\ref{fig:dj_standard_b} allows a graphically simpler connection to the  $|011\rangle$, $|100\rangle$, $|101\rangle$, and $|110\rangle$ states. For the sake of further simplifying transformations we will use the first representation.
	
	\subsection{Global Rotation and Target Severing}
	By applying the global Hadamard transformation required by the DJ algorithm, we rotate the operation from the X-basis (bit-flip) to the Z-basis (phase-flip). Because the ancilla qubit is initialized to the $|-\rangle$ state (the logical $|1\rangle_{F}$ in the rotated frame), the phase kickback mechanism acts as a static geometric symmetry rather than a dynamic flow of information. 
	
	\begin{figure}[h]
		\centering
		\includegraphics[width=0.8\linewidth]{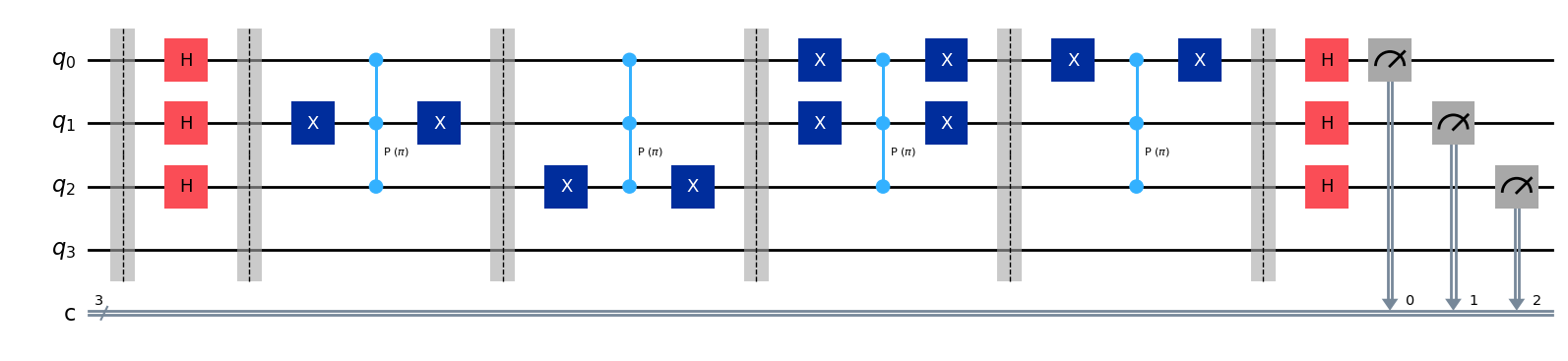}
		\caption{Factoring out the topological twist and collapsing the control states.}
		\label{fig:dj_step1__ancilla_severing}
	\end{figure}
	
	Applying the identity $H \cdot \text{MCX} \cdot H = \text{MCZ}$ converts the physical bit-flips into Multi-Controlled-Phase gates. Because the ancilla is locked in the excited phase state, its condition is perpetually satisfied. The ancilla line severs completely from the interaction (see Fig.~\ref{fig:dj_step1__ancilla_severing}), reducing the 4-qubit bit-flip oracle into a 3-qubit phase one. We will keep it for the sake of reference. Though it raises an immediate point about teaching the DJ algorithm using circuits without the ancilla qubit and only $MCZ$ gates.
	
	\subsection{The Algebraic Collapse of the Phase Polynomial}
\begin{figure}[h]
		\centering
		\begin{subfigure}[b]{0.48\linewidth}
			\centering
			\includegraphics[width=\linewidth]{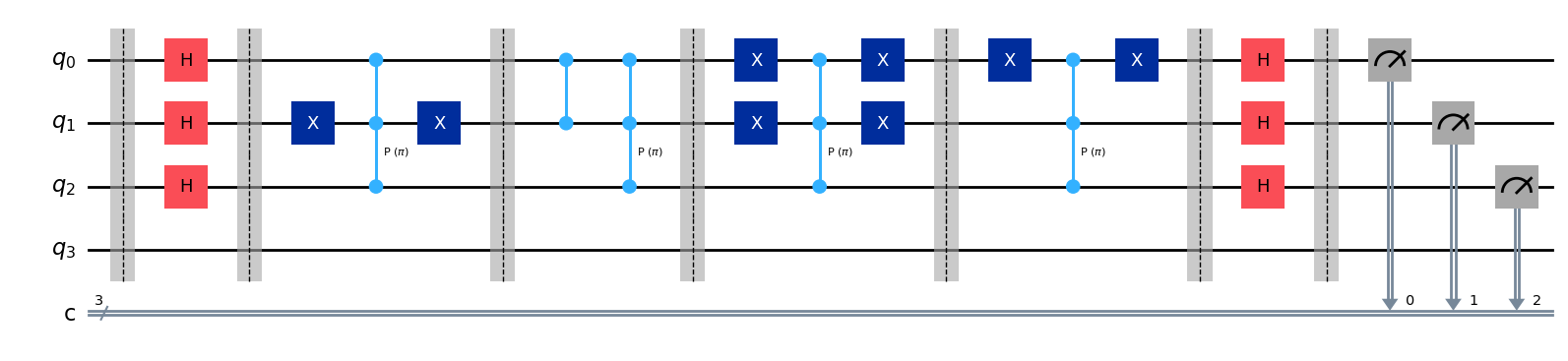}
			\caption{Pushing the X gate through the CCZ gate.}
			\label{fig:dj_step2_x_pushthrough}
		\end{subfigure}
		\hfill
		\begin{subfigure}[b]{0.48\linewidth}
			\centering
			\includegraphics[width=\linewidth]{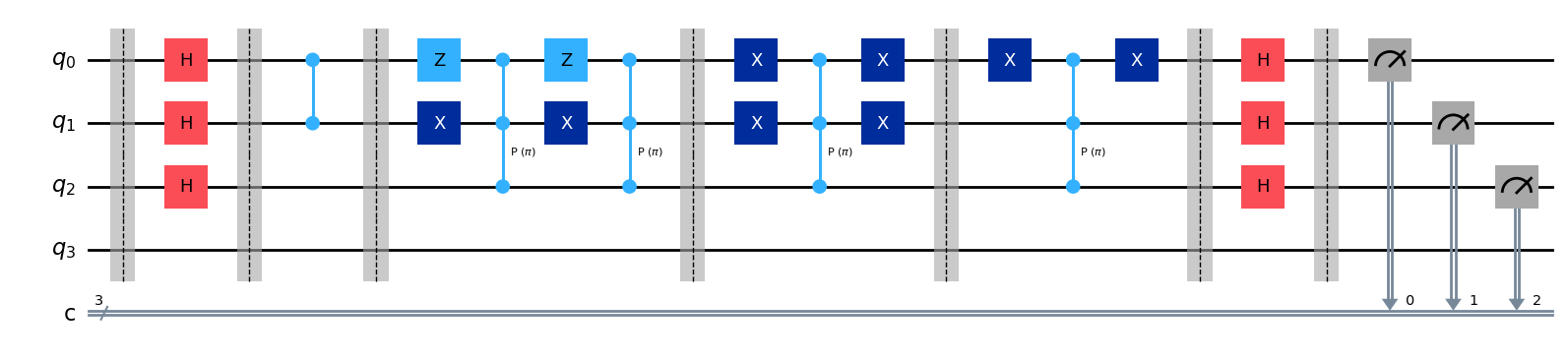}
			\caption{Pushing the CZ gate through the X and CCZ gates.}
			\label{fig:dj_step3_cz_pushthrough}
		\end{subfigure}
		
		\caption{The DJ oracle transformation by gate push-through.}
		\label{fig:dj_gate_pushthrough}
	\end{figure}

	Once in the Z-basis, the internal classical logic of the oracle is subjected to geometric simplification, proceeding in following phases:
	
	\textbf{1. Gate push-through:} For this particular case, we choose the $X$ on the $q_2$ line and push it through the $CCZ$ gate in the second string block, resulting in an $X X = I$ removal and a new $CZ$ gate - see Fig.~\ref{fig:dj_step2_x_pushthrough}. We follow by pushing the resultant $CZ$ gate to the leftmost side of the oracle, past the $X$ gates and past the $CCZ$ using its commutativity. This results in spawning two more $Z$ gates - see Fig.~\ref{fig:dj_step3_cz_pushthrough}. These $Z$ gates reduce via $Z Z = I$ leaving only $CCZ$ gates with accompanying $X$ gates. 
	An interesting topic to show students is the ``generation'' of twists (spawning gates) due to these transformations!
	
	\textbf{2. Geometric Exhaustion:} Factoring out the $CZ$ gate leaves behind four standard MCZ blocks on the $q_2$ target line, triggering on the control states $00$, $01$, $10$, and $11$. Because this exhaustively covers the entire Hilbert space of $q_0$ and $q_1$, the specific control conditions are rendered irrelevant. The basis states merge symmetrically ($\text{MCZ} \to \text{CZ} \to Z$), algebraically collapsing the four complex control blocks into a single, unconditional $Z$ gate on $q_2$.
	
	\begin{figure}[h]
		\centering
		\begin{subfigure}[b]{0.48\linewidth}
			\centering
			\includegraphics[width=\linewidth]{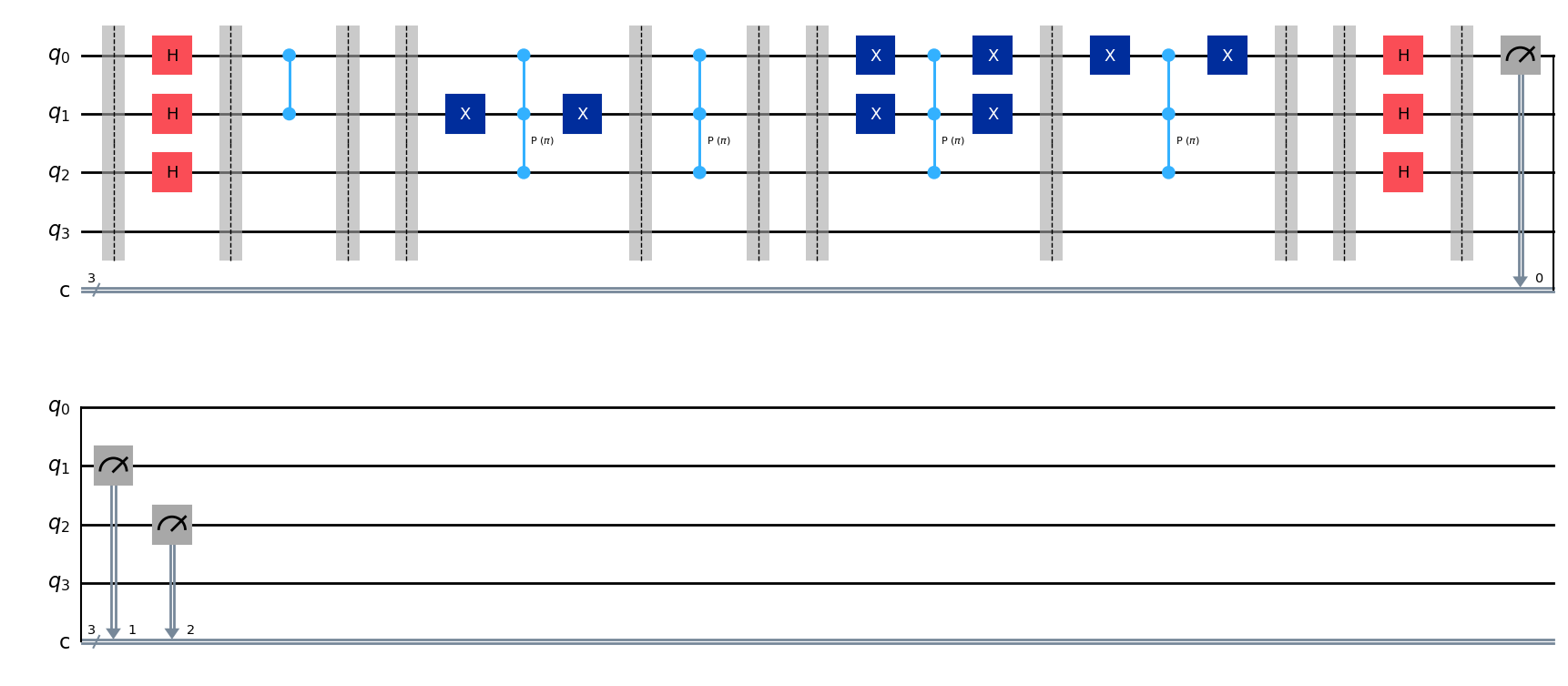}
			\caption{Grouping gates to reduce the q1 dependency.}
			\label{fig:dj_step4_q1_control_cut}
		\end{subfigure}
		\hfill
		\begin{subfigure}[b]{0.48\linewidth}
			\centering
			\includegraphics[width=\linewidth]{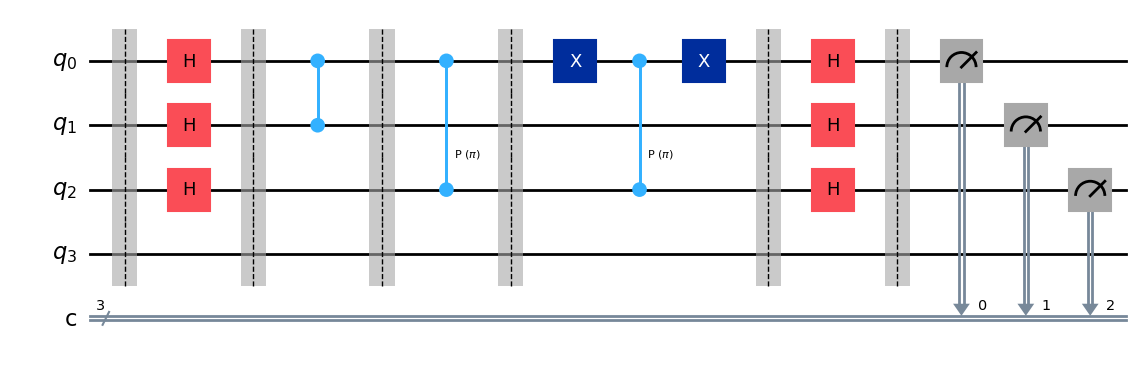}
			\caption{Grouping gates to reduce the q0 dependency.}
			\label{fig:dj_step5_q0_control_cut}
		\end{subfigure}
		
		\caption{The DJ oracle transformation by gate push-through.}
		\label{fig:dj_control_cuts}
	\end{figure}
	
	\subsection{The Irreducible Form and Future Work}
	The massive multi-controlled sequence is mechanically reduced to just two operations: a $CZ$ gate between $q_0$ and $q_1$, and a $Z$ gate on $q_2$. The representation of the Deutsch-Jozsa quantum oracle for this specific case is:
	\begin{equation}
		U_{\text{oracle}} = CZ_{0,1} \otimes Z_2
	\end{equation}
	
	Looking at the previous form, one can obviously see that the oracle can be globally rotated using the $H$ wrapper into:
	
	\begin{equation}
		U_{\text{oracle}} = RXX_{0,1} \cdot RX_{0} \otimes RX_{1} \otimes X_2
	\end{equation}

	\begin{figure}[h]
		\centering
		\begin{subfigure}[b]{0.48\linewidth}
			\centering
			\includegraphics[width=\linewidth]{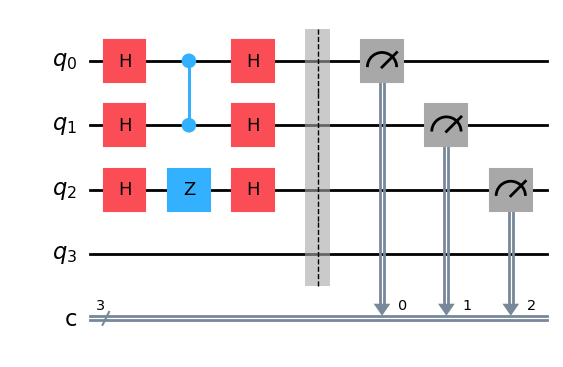}
			\caption{X-basis form}
			\label{fig:dj_final_1}
		\end{subfigure}
		\hfill
		\begin{subfigure}[b]{0.48\linewidth}
			\centering
			\includegraphics[width=\linewidth]{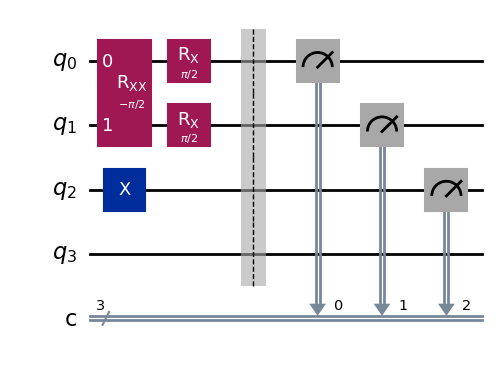}
			\caption{Z-basis form}
			\label{fig:dj_final_2}
		\end{subfigure}
		
		\caption{The final form of the quadratic oracle.}
		\label{fig:dj_final}
	\end{figure}
	
	What initially appeared to be a complex, highly entangled quantum search is revealed in the rotated basis to be a trivial evaluation of the phase polynomial $f(q) = (q_0 \cdot q_1) \oplus q_2$. 
	The somewhat ``mysterious'' quantum parallelism is more simply explained as classical algebra evaluated in the conjugate Fourier coordinate system, with the genuine quantum entanglement originating entirely from the non-aligned topological twist of the Mølmer–Sørensen $RXX$ gate. 
	While the Bernstein-Vazirani oracle (Family II) operates entirely as linear boolean algebra, the Deutsch-Jozsa oracle may introduce non-linearity (e.g., the quadratic term above).
	This is a further example of possible benefits of teaching about parity gates and the Cartan decomposition (see \cite{kraus2001, zhang2003}).
	Furthermore, by demonstrating the equivalence of circuits in both the X and Z bases, the students' attention may be redirected toward the truly indispensable quantum resource: entanglement.
	
	\vspace{1em}
	\noindent \textbf{Extensions to the Geometric Taxonomy:} 
	The geometric taxonomy reveals that quantum complexity is a direct manifestation of the oracle's algebraic degree over GF(2). Linear oracles (Degree 1) reduce purely to local $Z$ rotations, effectively recovering the Bernstein-Vazirani algorithm (Family II). Quadratic oracles (Degree 2, as shown above) introduce bipartite topological twists ($CZ$), generating Clifford entanglement (Family III). 
	
	A comprehensive formalization of this algebraic hierarchy---including the demonstration of how Degree 3 (cubic) algorithms generate tripartite phase twists ($CCZ$) that breach the Gottesman-Knill theorem to achieve universal computation---will be detailed in a forthcoming paper focusing specifically on the Deutsch-Jozsa algorithm. Furthermore, the geometric origin of entanglement through topological twisting in Family III will be explored in an upcoming analysis of Simon's algorithm.
	
	\vspace{1em}
	\noindent \textbf{Connections to Diagrammatic Formalisms:} 
	The geometric reduction demonstrated here shares a profound conceptual affinity with advanced diagrammatic languages such as the ZX and ZH-calculus (see Chapter 8 in ~\cite{vandewetering2020}). 
	In the ZH-calculus, boolean oracles are not represented by sequential Toffoli gates, but natively as multi-legged H-boxes (generalized multi-controlled Z-phase polynomials). For an undergraduate, the conceptual leap to this topological representation can be jarring. 
	By explicitly teaching the Hadamard wrapper as a geometric basis rotation that algebraically collapses bit-flip oracles into phase oracles, we provide the exact pedagogical scaffolding required to understand these modern formalisms. The geometric distinction between Family II and Family III circuits provides a natural stepping stone from standard quantum circuit models to advanced topological reasoning.
	
	\section*{Appendix B: Circuit Simulations in Qiskit}
	\label{app:qiskit_code}
	
	Below we provide the Qiskit implementations used to validate the geometric transformations discussed in the main text. 
	The code demonstrates the equivalence between the classical writing operation in the rotated frame and the standard Bernstein-Vazirani quantum circuit.
	
	To validate this geometric reframing, we implement the circuit in Qiskit in Jupyter Notebook format. 
	We demonstrate that the standard BV circuit produces identical statistics to the ``classical'' circuit in the rotated basis.
	
	\begin{lstlisting}[language=Python, caption={Bernstein-Vazirani in the X Basis}, label={lst:circuit_1_bv_xwrite}]
		# @title BV in the X basis (step 1)
		
		def create_bv_x1_circuit(secret_string):
		"""
		Creates a simple classical 'WRITE-SECRET' circuit using CX gates and the X ancilla prep.
		"""
		n = len(secret_string)
		# n data qubits + 1 ancilla (the last qubit)
		qc = QuantumCircuit(n + 1, n)
		
		# --- 0. PREPARE ANCILLA ---
		ancilla_index = n
		qc.barrier()
		
		# --- 1. INITIALIZATION ---
		# '1' on ancilla means: WRITE THE SECRET
		qc.x(ancilla_index)
		qc.barrier()
		
		# --- 2. THE 'ORACLE' (CX GATES) ---
		# We iterate through the secret string.
		# If the ancilla qubit is '1', we write the string into the measured qubits
		s_rev = secret_string[::-1]
		
		for qubit_index, bit in enumerate(s_rev):
		if bit == '1':
		qc.cx(ancilla_index, qubit_index)
		
		qc.barrier()
		
		# --- 4. MEASUREMENT ---
		# Verify the bits are written as requested
		qc.measure(range(n), range(n))
		
		return qc
		
		
		# @title BV in the X basis (step 1)
		# --- CONFIGURATION ---
		secret_s = "10110011"
		circuit = create_bv_x1_circuit(secret_s)
		
		# --- VISUALIZATION ---
		fig = circuit.draw('mpl', style='iqp', scale=1.3, filename='bv_x_circuit__2.png')
		
		# Display the diagram
		print(f"Generating Circuit for Secret String: {secret_s}")
		display(fig)
		
		# --- SIMULATION PROOF ---
		# Let's run it to prove your post is correct!
		simulator = AerSimulator()
		compiled_circuit_x = transpile(circuit, simulator)
		result_x = simulator.run(compiled_circuit_x, shots=1024).result()
		counts_x = result_x.get_counts()
		
		print("\n--- SIMULATION RESULTS ---")
		print(f"Secret String: {secret_s}")
		print(f"Measurement:   {counts_x}")
		print("Did it work?   ", "YES" if list(counts_x.keys())[0] == secret_s else "NO")
	\end{lstlisting}
	
	\begin{lstlisting}[language=Python, caption={Bernstein-Vazirani in the Z Basis with CZ oracle}, label={lst:circuit_2_bv_zcz}]	
		def create_bv_x2c_circuit(secret_string):
		"""
		Creates a Bernstein-Vazirani circuit using H CZ H gates and the X ancilla prep.
		"""
		# n data qubits + 1 ancilla (the last qubit)
		n = len(secret_string)
		qc = QuantumCircuit(n + 1, n)
		
		# Note: We reverse the string to match Qiskit's qubit ordering (q0 is rightmost)
		s_rev = secret_string[::-1]
		
		# --- 0. PREPARE ANCILLA ---
		ancilla_index = n
		qc.barrier()
		
		# --- 1. INITIALIZATION ---
		qc.x(ancilla_index)
		qc.barrier()
		
		# We fill the missing Hadamard gates
		qc.h(range(n))
		qc.barrier()
		
		# --- 2. THE ORACLE (CZ GATES) ---
		# We iterate through the secret string.
		# If the bit is '1', we perform a CZ between that data qubit and the ancilla.
		#
		# The CZ transforms |+> and |-> states
		# and Hadamard rotates the X-basis states into the Z-basis output
		for qubit_index, bit in enumerate(s_rev):
		if bit == '1':
		qc.cz(ancilla_index, qubit_index)
		qc.barrier()
		
		# We fill the missing Hadamard gates
		qc.h(range(n))
		qc.barrier()
		
		# --- 4. MEASUREMENT ---
		# Measure data qubits into classical bits
		qc.measure(range(n), range(n))
		
		return qc
		
		
		# @title BV in the Z basis CZ oracle
		# --- CONFIGURATION ---
		secret_s = "10110011"
		circuit = create_bv_x2c_circuit(secret_s)
		
		# --- VISUALIZATION FOR LINKEDIN ---
		fig = circuit.draw('mpl', style='iqp', scale=1.3, filename='bv_x_circuit__2c.png')
		
		# Display the diagram
		print(f"Generating Circuit for Secret String: {secret_s}")
		display(fig)
		
		# --- SIMULATION PROOF ---
		simulator = AerSimulator()
		compiled_circuit_x_2c = transpile(circuit, simulator)
		result_x_2c = simulator.run(compiled_circuit_x_2c, shots=1024).result()
		counts_x_2c = result_x_2c.get_counts()
		
		print("\n--- SIMULATION RESULTS ---")
		print(f"Secret String: {secret_s}")
		print(f"Measurement:   {counts_x}")
		print("Did it work?   ", "YES" if list(counts_x.keys())[0] == secret_s else "NO")
	\end{lstlisting}
	
	\begin{lstlisting}[language=Python, caption={Bernstein-Vazirani in the Z Basis with CX oracle}, label={lst:circuit_3_bv_zcx}]
		
		# @title BV in the X basis (FINAL)
		
		def create_bv_x_final_circuit(secret_string):
		"""
		Creates a 'regular' Bernstein-Vazirani circuit using CX gates and the X ancilla prep.
		"""
		n = len(secret_string)
		# n data qubits + 1 ancilla (the last qubit)
		qc = QuantumCircuit(n + 1, n)
		
		# --- 0. PREPARE ANCILLA ---
		ancilla_index = n
		qc.barrier()
		
		# --- 1. INITIALIZATION ---
		# Prepare ancilla qubit
		qc.x(ancilla_index)
		qc.barrier()
		
		# Apply H to all data qubits
		qc.h(range(n+1))
		qc.barrier()
		
		# --- 2. THE ORACLE (CX GATES) ---
		# We iterate through the secret string.
		# If the ancilla qubit is '1', the CX kicks the phase back to measured qubits
		# and Hadamard rotates the phase into the Z-basis output
		s_rev = secret_string[::-1]
		
		for qubit_index, bit in enumerate(s_rev):
		if bit == '1':
		qc.cx(qubit_index, ancilla_index)
		qc.barrier()
		
		# --- 3. MEASUREMENT BASIS CHANGE ---
		# Apply H to data qubits to switch back from X-basis to Z-basis
		qc.h(range(n+1))
		qc.barrier()
		
		# --- 4. MEASUREMENT ---
		# Measure data qubits into classical bits
		qc.measure(range(n), range(n))
		
		return qc
		
		
		# @title BV in the X basis (FINAL)
		# --- CONFIGURATION ---
		secret_s = "10110011"
		circuit = create_bv_x_final_circuit(secret_s)
		
		# --- VISUALIZATION ---
		fig = circuit.draw('mpl', style='iqp', scale=1.3, filename='bv_x_circuit__1.png')
		
		# Display the diagram
		print(f"Generating Circuit for Secret String: {secret_s}")
		display(fig)
		
		# --- SIMULATION PROOF ---
		simulator = AerSimulator()
		compiled_circuit_x_final = transpile(circuit, simulator)
		result_x_final = simulator.run(compiled_circuit_x_final, shots=1024).result()
		counts_x_final = result_x_final.get_counts()
		
		print("\n--- SIMULATION RESULTS ---")
		print(f"Secret String: {secret_s}")
		print(f"Measurement:   {counts_x}")
		print("Did it work?   ", "YES" if list(counts_x.keys())[0] == secret_s else "NO")
	\end{lstlisting}

\end{document}